\documentclass[useAMS,usenatbib]{mn2e}
\usepackage[dvips]{epsfig}

\def\aap{\rm{A\&A}}                
\def\apj{\rm{ApJ}}                 
\def\apjl{\rm{ApJ}}                
\def\apjs{\rm{ApJS}}               
\def\mnras{\rm{MNRAS}}             
\def\prd{\rm{Phys.~Rev.~D}}        

\begin{document}

\title[Non-Gaussian Foreground Residuals]{Non-Gaussian Foreground
  Residuals of the WMAP First Year Maps} \author[Jo\~{a}o Medeiros and Carlo
R. Contaldi]{Jo\~{a}o Medeiros\thanks{joao.medeiros@imperial.ac.uk}
and Carlo R. Contaldi \\Theoretical Physics, Imperial College, Prince
Consort Road, London SW7 2BZ, UK}

\date{\today}

\maketitle
\begin{abstract}
We investigate the effect of foreground residuals in the WMAP data
\citep{wmap} by adding foreground contamination to Gaussian ensembles
of CMB signal and noise maps. We evaluate a set of non-Gaussian
estimators on the contaminated ensembles to determine with what
accuracy any residual in the data can be constrained using higher
order statistics. We apply the estimators to the raw and cleaned Q, V,
and W band first year maps. The foreground subtraction method applied to
{\sl clean} the data in \citet{wmapfor} appears to have induced a correlation
between the power spectra and normalized bispectra of the maps which
is absent in Gaussian simulations. It also appears to increase the
correlation between the $\Delta\ell=1$ inter-$\ell$ bispectrum of the
cleaned maps and the foreground templates. In a number of cases the
significance of the effect is above the 98\% confidence level.
\end{abstract}


\section{Introduction}\label{introduction}

Over the past few years there has been a heightened interest in
testing the statistical properties of Cosmic Microwave Background
(CMB) data. The process has been accelerated by the release of the
WMAP first year results. The WMAP data provide the first ever,
full-sky maps which are signal dominated up to scales of a few
degrees. Thus, for the first time we can test the Gaussianity and
isotropy assumptions of the cosmological signal over large scales in
the sample variance limit.

Ever since the release of the COBE-DMR results \citep{cobe} a
consensus has been hard to reach on tests of non-Gaussianity with some
studies reporting null results \citep{kog96a,contaldi2000,sandvik}
while others claimed detections of non-Gaussian features
\citep{fmg,joao,novikov, pando}. With the release of the WMAP first
year results a limit on the non-Gaussianity of primordial
perturbations in the form of an estimate of the non-linear factor
$f_{\rm NL}$ was obtained by \cite{wmapKomatsu}. However a number of
authors \citep{hbg,erik1,coles,park,copi,teglarge,kate,efstathiou,
football,npoint,jaffe} have also reported analysis of the maps that
suggest violations of the Gaussian or isotropic nature of the signal.

One of problems with testing Gaussianity is that one can devise a
plethora of tests to probe the infinite degrees of non-Gaussianity,
therefore different tests represent different perspectives on the
statistical patterns of the signal. For WMAP there are already a
number of detections of so called anomalies, most pointing to
different unexpected features in the microwave sky. The most
documented case \citep{peiris,efstathiou,slos,teglarge,erik1}
is the low amplitude of the quadrupole and octupole in comparison to
the inflationary prediction, something we can categorize as
{\sl amplitude anomalies}. Although it is simple to design inflationary
spectra with sharp features which reproduce, more or less closely, the
amplitude anomaly (see e.g. \cite{contaldi2003,bridle,salopek}) these
invariably suffer fine tuning problems. Another approach is to relate
the anomaly to the breakdown of statistical isotropy or Gaussianity.

Other reported features relate to the correlation of phases in the
multipole coefficients which are an indication of
non-Gaussianity. These can be dubbed {\sl phase anomalies}. One
example is the hemisphere asymmetries \citep{erik1}; the northern
ecliptic hemisphere is practically flat while the southern hemisphere
displays relatively high fluctuations in the power spectrum.  Other
functions, such as the bispectrum \citep{kate} and n-point correlation
functions \citep{npoint} also show related asymmetries. Furthermore,
there is the anomalous morphology of several multipoles, in
particular, the striking planarity of the quadrupole and octupole and
the strong alignment between their preferred directions
\citep{teglarge}. Overall, there is a strong motivation to continue
probing the statistical properties of the data and find possible
sources for these signals, be it instrumental, astrophysical or
cosmological.

The first test to have provided indications of possible non-Gaussian
features in the CMB data was reported by \cite{fmg} and \cite{joao}
using a bispectrum estimator, the fourier analog of the three point
function. Both those detections were later found to be caused by
systematic effects rather than by cosmological source as reported by
\cite{banday}. For the case of the bispectrum signal detected by
\cite{joao}, which used an estimator tuned to detect correlations
between neighbouring angular scales, finding the source of the signal
had to wait for the release of the high precision WMAP data
\citep{wmap} which was able to provide a comparative test of the
cosmological signal. The WMAP data did not reproduce COBE's result and
systematic errors were found to be the cause \citep{joaoes}. The WMAP
data was later analysed with the bispectrum in more detail
by \cite{kate}. In that paper, the bispectrum of the clean, coadded maps
was analysed and a connection between the hemisphere asymmetries in
the 3-point correlation function and the bispectrum was established,
although the full sky as a whole was found to be consistent with
Gaussianity.

In this paper, we study the effect that foreground contaminations have
on bispectrum estimators. In section~\ref{sec:def} we define a
set of bispectrum estimators with set $\ell$ configurations. In
section~\ref{sec:foregrounds} we describe the template dust, free-free
and synchrotron maps used to characterize the effect on the
bispectrum. In section~\ref{sec:method} we determine the distribution
of the estimators in the presence of residual foregrounds with
different amplitudes and discuss the application of this method to
detect residuals in the data by introducing a number of statistical
and correlation measures. In section~\ref{sec:application} we discuss
the application of the the statistical tools developed in the previous
sections to the raw and cleaned WMAP first year maps. We conclude with
a discussion of our method and results in section~\ref{sec:disc}.

\section{The Angular Bispectrum}\label{sec:def}

We now introduce the angular bispectrum estimator \citep{fmg}. The
bispectrum is related to the third order moment of the spherical
harmonic coefficients $a_{\ell m}$ of a temperature fluctuation map
$\Delta T({\bf \hat n})/T$. The coefficients describe the usual
expansion of the map over the set of spherical harmonics $Y_{\ell
m}({\bf \hat n})$ as
\begin{eqnarray}
\frac{\Delta T}{T}({\bf \hat n})=
\sum_{\ell m}a_{\ell m}Y_{\ell m}({\bf \hat n}).
\end{eqnarray}

Given a map, either in pixel space or harmonic space, and assuming
statistical isotropy, one can construct a set hierarchy of
rotationally invariant statistical quantities characterizing the
pattern of fluctuations in the maps. These are the n-point correlation
functions in the temperature fluctuations $\langle\frac{\Delta
T}{T}({\bf \hat m})\frac{\Delta T}{T}({\bf \hat n})...\frac{\Delta
T}{T}({\bf \hat p})\rangle$ or in the spherical harmonic coefficients,
$\langle a_{\ell_1 m_1}a_{\ell_2m_2}...a_{\ell_nm_n}\rangle$. 

The unique {\it quadratic} invariant is the angular power spectrum
defined as
$\langle a_{\ell_1 m_1} a^\star_{\ell_2 m_2}\rangle = \delta_{\ell_1\ell_2}\delta_{m_1m_2}C_\ell$, whose estimator can be written as ${\hat
C}_\ell=\frac{1}{2\ell+1}\sum_m|a_{\ell m}|^2$.  This gives a measure
of the overall intensity for each multipole $\ell$.  Following
\cite{fmg}, the most general {\it cubic} invariant defines the angle
averaged bispectrum,
\begin{equation}
\left<a_{\ell_1 m_1}a_{\ell_2 m_2}a_{\ell_3 m_3}\right>=
B_{\ell_1\ell_2\ell_3}\left (
\begin{array}{ccc} \ell_1 & \ell_2 & \ell_3 \\ m_1 & m_2 & m_3
\end{array} \right ),
\end{equation}
where the $(\ldots)$ is the Wigner $3J$ symbol. Parity invariance of
the spherical harmonic functions dictates that the bispectrum be
non-zero only for multipole combinations where the sum
$\ell_1+\ell_2+\ell_3$ is even.  An unbiased estimator (for the full
sky) can be evaluated as
\begin{eqnarray}
{\hat B}_{\ell_1\ell_2\ell_3}&=&\frac{{\cal N}^{-1}_{\ell_1\ell_2\ell_3}}{\sqrt{4\pi}}\sum_{m_1m_2m_3}\left ( \begin{array}{ccc} \ell_1 & \ell_2 & \ell_3
\\ m_1 & m_2 & m_3
\end{array} \right )\times\\ 
&&a_{\ell_1 m_1}a_{\ell_2 m_2} a_{\ell_3 m_3},\nonumber
\end{eqnarray}
with the normalization factor defined as
\begin{eqnarray}
{\cal N}_{\ell_1\ell_2\ell_3}&=&{\left
(\begin{array}{ccc} \ell_1 & \ell_2 & \ell_3 \\ 0 & 0 & 0\end{array}
\right
)}\times\\&&\sqrt{\frac{(2\ell_1+1)(2\ell_2+1)(2\ell_3+1)}{4\pi}}.\nonumber
\end{eqnarray}
The bispectrum can be related to the three-point correlation functions
of the map just as the power spectrum $C_\ell$ can be related to the
correlation function $C(\theta)$ through the well known expression
\begin{equation}
C(\theta) = \frac{1}{4\pi}\sum_\ell (2\ell+1)C_\ell P_\ell(\cos\theta).
\end{equation}
For example, the pseudo-collapsed,
three-point correlation function, $C^{(3)}(\theta)=\langle
\frac{\Delta T}{T}({\bf \hat n})^2\frac{\Delta T}{T}({\bf \hat m})
\rangle$, is related to our definition of the bispectrum $
B_{\ell_1\ell_2\ell_3}$ as
\begin{equation}\label{3pt}
C^{(3)}(\theta)= \frac{1}{4\pi}\sum_{\ell_1\ell_2\ell_3}{\cal N}_{\ell_1\ell_2\ell_3}B_{\ell_1\ell_2\ell_3}P_{\ell_3}(\cos\theta),
\end{equation}
where ${\bf \hat n}\cdot{\bf \hat m}=\cos\theta$.

It is important to use both tools, the bispectrum and the three-point
correlation function, to probe the sky maps as they have the capacity
to highlight different features of the data. In principle, harmonic
space based methods are preferred for the study of primordial
fluctuations whereas real space methods are more sensitive to
systematics and foregrounds, which are strongly localized in real
space. In addition, the three-point correlation function is intrinsically very
sensitive to the low-$\ell$ modes, whereas the bispectrum can pick up
different degrees of freedom with respect to the different mode
correlations we want to probe

For the choice $\ell_1=\ell_2=\ell_3=\ell$ we can define the
single-$\ell$ bispectrum ${\hat B_\ell}=\hat B_{\ell\, \ell \,\ell}$
\citep{fmg}, which probes correlations between different $m$'s.  Other
bispectrum components are sensitive to correlations between different
scales $\ell$. This can be extended to study correlations from
different angular scales. The simplest of these is the $\Delta\ell=1$
inter-$\ell$ bispectrum between neighbouring multipoles defined as
$\hat B_{\ell-1\, \ell \,\ell+1}$ \citep{joao}. It is convenient to
consider estimators normalized by their expected Gaussian variance
$\hat C_{\ell_1}\hat C_{\ell_2}\hat C_{\ell_3}$ which have been shown
to be more optimal and more Gaussian distributed than the unnormalized
estimators, and are not sensitive to the overall power in the
maps. Here we will introduce the $\hat I_\ell$,$\hat J_\ell$, and
$\hat K_\ell$ bispectra defined as
\begin{equation}\label{i3j3}
I^3_\ell = { {\hat B}_{\ell} \over ({\hat C}_{\ell})^{3/2}} , \ \
J^3_\ell = { \hat B_{\ell-1\, \ell \,\ell+1} \over ({\hat C}_{\ell-1}{\hat
C}_{\ell} {\hat C}_{\ell+1})^{1/2}},
\end{equation}
and
\begin{equation}\label{k3}
K^3_\ell = {  \hat B_{\ell-2\, \ell \,\ell+2}
\over ({\hat C}_{\ell-2}{\hat C}_{\ell}
{\hat C}_{\ell+2})^{1/2}},
\end{equation}
where have extended the formalism to a separation $\Delta\ell=2$ to
probe signals with both odd and even parity in the inter-$\ell$
correlations.

\section{Foreground Templates}\label{sec:foregrounds}

The standard method of foreground removal used by cosmologists
makes use of a set of template maps for each of the dominant
sources of foreground contamination in the CMB frequency maps.
These are maps obtained from independent astronomical full-sky
observations at frequencies where the respective mechanisms of
emission are supposed to be dominant. These templates are the H
$\alpha$ map \citep{halpha}, for the free-free emission, the 408
MHz Haslam map \citep{haslam}, for the synchrotron emission, and
the FDS 94 GHz dust map \citep{FDS}. These are then subtracted
from the WMAP data with coupling coefficients determined by cross
correlating with the observed maps in the Q (41 GHz), V (61 GHz),
and W (94 GHz) bands. Nevertheless the templates are a poor
approximation of the of the real sky near the galactic plane, so a
Kp2 mask must still be used in the analysis. The method is
described in \cite{wmapfor} and \cite{komatsu};
\begin{eqnarray}\label{eq:amp}
\overline{T}_{Q} &=& T_{Q} - 1.044\,[1.036\, T^{\rm FDS} +
 \frac{1.923}{\eta}\,T^{H\alpha} \nonumber\\ &&+1.006\,T^{\rm
 Sync}],\nonumber\\ \overline{T}_{V} &=& T_{V} - 1.100\,[0.619\,
 T^{\rm FDS} + \frac{1.923}{\eta} \,\left(\frac{\nu_{V}}{
 \nu_{Q}}\right)^{-2.15}\,T^{H\alpha} \nonumber\\ && +1.006
 \,\left(\frac{\nu_{V}}{ \nu_{Q}}\right)^{-2.7}\,T^{\rm Sync}],\\
 \overline{T}_{W} &=& T_{W}- 1.251[0.873\, T^{\rm FDS} +
 \frac{1.923}{\eta}\, \left(\frac{\nu_{W}}{
 \nu_{Q}}\right)^{-2.15}\,T^{H\alpha} \nonumber \\
 &&+1.006\,\left(\frac{\nu_{W}}{ \nu_{Q}}\right)^{-2.7}\,T^{\rm
 Sync}],\nonumber
\end{eqnarray}
where $ \eta$ is a correction factor due to reddening in the free-free
template and $\nu_{Q}=40.7$ GHz, $\nu_{Q}=60.8$ GHz and $\nu_{W}=93.5$
GHz . The values in front of the left bracket convert the detector's
temperature to thermodynamic temperature.  It is considered that this
is a sufficiently good method to remove the foregrounds outside the
Kp2 plane since it matches the correct amplitudes quite well, however
the usual doubts remain, especially in the light of the alignment/low
multipoles controversies. Another point one can make is that
whereas this may be a satisfactory technique to correct the
foregrounds at the power spectrum level, its effect on higher order
statistics is unknown and may actually induce unexpected correlations.

\section{The effect of foregrounds on the bispectrum}\label{sec:method}

\begin{figure*}
\centerline{\psfig{file=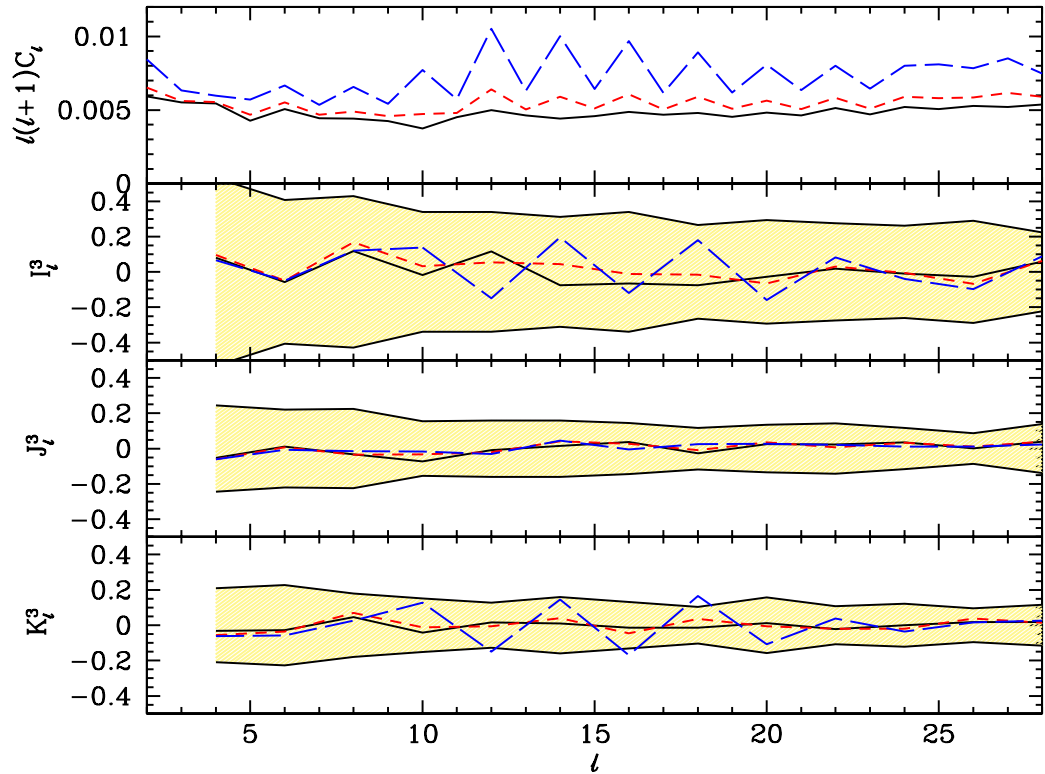,angle=270,width=15cm}}
 \caption{ The average functions for the angular spectra of the
 simulations. Black (solid) is Gaussian, red (short-dashed) is for the
 contaminated simulations with $\alpha=0.5$ and blue (long-dashed) is
 for the contaminated simulations with $\alpha=1.0$.  The top panel
 shows the power spectrum $\hat C_\ell$, second panel shows the
 single-$\ell$ bispectrum $I^3_\ell$, the third panel the
 $\Delta\ell=1$ inter-$\ell$ bispectrum $J^3_\ell$ and the bottom
 panel shows the $\Delta\ell=2$ inter-$\ell$ bispectrum
 $K^3_\ell$. The shaded regions represent the Gaussian variance
 measured directly from the ensemble of simulations.  For the Gaussian
 simulations, the average power spectrum is just the input $\Lambda$CDM
 power spectrum and the average bispectra is effectively zero. On the
 other hand, the average angular spectra of the contaminated
 simulations have an emerging pattern of intermittency in both second-
 and third-order statistics. This intermittent pattern comes about
 due to the even parity of galactic foregrounds, ie, even modes
 are enhanced relatively to the odd modes. This can be seen in the
 significant increase of power in the even modes of the power
 spectrum. In terms of the bispectrum, we see that the $\Delta\ell=0$
 and the $\Delta\ell=2$ inter-$\ell$ components will be more
 significantly enhanced than the $\Delta\ell=1$ inter-$\ell$
 bispectrum because the latter includes correlations between even and
 odd modes.}\label{fig:bisp}
\end{figure*}

We have generated a set of 3000 Gaussian, CMB simulations of the WMAP
first year Q, V, and W maps in
\textsc{HEALPix}\footnote{http://healpix.jpl.nasa.gov}\citep{healpix}
format with a resolution parameter $N_{\rm side}=512$. Each simulation
is smoothed with the Q, V and W frequency channel beams and channel
specific noise is added. We adopted the WMAP best-fit $\Lambda$CDM
with running index power spectrum\footnote{http://lambda.gsfc.nasa.gov} to generate the $a_{\ell m}$
coefficients of the maps. The Kp2 galactic mask is imposed on each
map. The masked maps are then decomposed into spherical harmonic
coefficients $a_{\ell m}$ using the \textsc{Anafast} routine. We then
calculate the four spectra; namely the the power spectrum $\hat
C_\ell$, single-$\ell$ bispectrum $I^3_\ell$, $\Delta\ell=1$
inter-$\ell$ bispectrum $J^3_\ell$ and $\Delta\ell=2$ inter-$\ell$
bispectrum $K^3_\ell$ as described in section~\ref{sec:def}.

We then add channel-specific foregrounds outside the Kp2 zone to the
same set of Gaussian simulations with amplitudes set as in
Eqn.~(\ref{eq:amp}). The addition of the foreground is scaled linearly
by a factor $\alpha$ as
\begin{equation}
T_{\rm \{Q,V,W\}} = T_{\rm CMB} + \alpha\overline T_{\rm \{Q,V,W\}},
\end{equation}
which we use to check the sensitivity of the bispectra to the
foregrounds (typically $\alpha=1.0 $ or $ \alpha=0.5$). The power
spectrum and bispectra are then calculated for the set of
contaminated maps.

In Fig.~\ref{fig:bisp} we show the mean angular spectra of the
simulations obtained by averaging over the ensembles. We show the mean
spectra for the Gaussian (solid, black) and the contaminated
simulations for $\alpha=0.5$ (short-dashed, red) and $\alpha=1.0$
(long-dashed, blue). The shaded area shows the variance of the three
bispectra obtained directly from the Gaussian simulations.

We see that even for the fully contaminated set of maps ($\alpha=1.0$)
the average signal is not significantly larger than the expected
Gaussian variance indicating that a detection would require averaging
over a large number of modes. However we see some important
distinguishing features in the signal in that it is sensitive to the
parity of the multipole, being suppressed for odd $\ell$. This is due
to the approximate symmetry of the foreground emission about the
galactic plane which means that most of the signal will be in even
$\ell$ modes since these have the same symmetry. This effect can be
seen in all the spectra but most significant is the suppression of the
odd inter-$\ell$ bispectrum $J^3_{\ell}$ with respect to the even
inter-$\ell$ bispectra $I^3_{\ell}$ and $K^3_{\ell}$.

Another obvious feature of the even parity  nature of the signal
is the correlation between the spectra. In particular the absolute
values of the $I^3_\ell$ and the $K^3_\ell$ are correlated with
the structure visible in the fully contaminated power
spectrum.

Overall the $K^3_\ell$ is the most sensitive statistic with the
largest amplitude with respect to the Gaussian variance although
still quite small even at $50\%$ contamination. We now describe a
number of statistical estimators we use to test the detectability
of the template matched foregrounds in the Q, V, and W channel
maps.

\subsection{Chi-Squared Test}\label{sec:chisq}

Having seen how foregrounds affect the angular statistics of CMB
maps, we can now devise specific tests to probe these properties on
the bispectrum and test their sensitivity.  The standard way to use
the bispectrum as a test of general non-Gaussianity is to use a
reduced $\chi ^2 $ statistic \citep{joao,joaoes,kate}. This is defined
as
\begin{equation}
\chi^2={1\over N_\ell}{\sum_{\ell = \ell_{\rm min}}^{\ell_{\rm max}}
\chi_\ell^2}= {1\over N_\ell} \sum_{\ell = \ell_{\rm min}}^{\ell_{\rm
max}} { {( X_\ell-\langle  X_\ell\rangle)^2} \over
{\sigma_\ell^2} }.
\end{equation}
where $ X_{\ell} $ is a given bispectrum statistic, $\langle
X_\ell\rangle$ is its mean value computed over the Monte Carlo
ensembles, and $\sigma_\ell^2$ is the variance for each angular
scale. The $\chi^2 $ test is a measure of the deviation of the
observed data from the expected mean, weighted by the Gaussian
variance of the estimator.

Foregrounds increase the amplitude of the bispectra foregrounds, but
as shown in Fig.~\ref{fig:bisp}, we can see that only $K^3_\ell$ seems
to stand of chance of significant detections since the average
amplitude of the signal is comparable to the variance, unlike the
other components of the bispectrum.

The detectability of the template matched signals using any of the
bispectra can be tested by comparing the distribution of the $\chi
^2 $ values obtained from the contaminated simulations with that
obtained from Gaussian simulations. We compute the $\chi^2 $
values for the contaminated maps using the mean and the variance
obtained from the Gaussian simulations, ie, the expected Gaussian
functions. 

We compare the distribution of the $\chi^2 $ values for the Gaussian
simulations against the distribution obtained for the simulations with
contamination ($\alpha=1.0$). We concentrate on the Q band since it is
the most contaminated frequency. The histograms of the $\chi^2$ are
shown in the left column of Fig.~\ref{fig:histo}. For the $I^3_\ell$
and $J^3_\ell$ spectra the histograms overlap completely. This means
that the probability of finding contaminated simulations with a high
$\chi^2$ is the same as for the Gaussian simulations indicating that
the $\chi ^2 $ test is insensitive to the presence of foreground
contaminations at this level. However the $K^3_\ell$ spectrum tells a
different story. There is a significant shift between the two
distributions which implies that this component of the bispectrum has
more sensitivity to foregrounds.

The sensitivity can be quantified in terms of the fraction of the
contaminated simulations (with $\alpha=1.$) with a $\chi^2$ larger
value than 95.45 $\%$ (i.e. 2 $\sigma$) of the Gaussian simulations
(with $\alpha =0$). The sensitivity for the $I^3_\ell$ and $J^3_\ell$
is low (< 0.05), whereas for $K^3_\ell$ the fraction increases to
0.355.

\subsection{Template Correlation Test}\label{sec:correlation}

A template matched statistic can be defined by correlating the
observed bispectra in the data with those of the foreground
templates. This is more sensitive to the structure in the template
signal as opposed to the $\chi^2$ test introduced above. We define
a cross correlation statistic $\rho$ as
\begin{equation}\label{corr_coef}
\rho=\frac{\sum_{\ell = \ell_{\rm min}}^{\ell_{\rm max}}{ X_{\ell} X^{F}_{\ell} }}{\left(\sum_{\ell = \ell_{\rm
min}}^{\ell_{\rm max}} X_{\ell}^2\sum_{\ell = \ell_{\rm
min}}^{\ell_{\rm max}} X^{F\,2}_{\ell}\right)^{1/2}}
\end{equation}
where $X_{\ell}$ are the bispectra obtained from the data and
the $X_{\ell}^{F}$ are those obtained from the foreground
templates.

In the middle column of Fig.~\ref{fig:histo} we display the histograms
for the $\rho $ values for the Gaussian simulations against the
distribution obtained for the contaminated ($\alpha=1$) simulations of
the Q band maps. The sensitivity has improved over the $\chi^2 $ test,
with the histograms of the input and output data sets being clearly
shifted, meaning that there is a higher probability of detection of
foregrounds using this method. Again the effect is stronger in the
$K^3_\ell$. This result simply quantifies the statement that a matched
template search for a contamination signal is more sensitive than a `blind'
statistic such as the $\chi^2$ test. The values for the sensitivity of the
test are given in table~\ref{tab:corr} for all three WMAP bands.

\subsection{Power Spectrum and Bispectra Cross-Correlation Test}\label{sec:rstat}

For a Gaussian field, the normalized bispectrum is statistically
uncorrelated with the power spectrum \citep{conf}. However,
foreground residuals in the map induce non-Gaussian correlations which
in turn will induce correlations between the normalized bispectra and
the power spectrum of the maps. This can provide another specific
signature that one can use to detect the presence of foreground
contamination.

For Gaussian simulations, the average power spectrum
is just the input $\Lambda$CDM power spectrum and the bispectrum is
effectively zero. On the other hand, the average angular spectra of
the contaminated simulations have an emerging pattern of intermittency
in both first- and second-order statistics. Correlations between the
power spectrum and the bispectra therefore come about due to the even
parity induced by the characteristic galactic foregrounds. This means
that the even modes of the power spectrum will be correlated
with the even modes of the bispectra, whereas odd modes will remain
uncorrelated. In order to test this effect on the maps, we introduce
the ${\rm R}$ correlation statistic defined as
\begin{equation}
{\rm R}^X = \sum^{\ell_{\rm max}}_{\ell=\ell_{\rm min}}(-1)^{{\rm
int}[\frac{\ell}{2}]+1} \hat C_\ell X_\ell
\end{equation}
where $\hat C_\ell$ is the observed power spectrum. We have chosen
$\ell_{\rm max}=30$ and $\ell_{\rm min}=4$ as we are interested in the
large angular scales where the effects of foreground contamination
will dominate. We use the absolute value of the bispectrum in order to
avoid the discrimination between negative and positive correlations
which would affect our sum. We are only interested in the
discrimination between the existence of absolute correlations against
null correlations between the $\hat C_\ell $ and the bispectra $X_\ell$.

\begin{figure*}
\centerline{\psfig{file=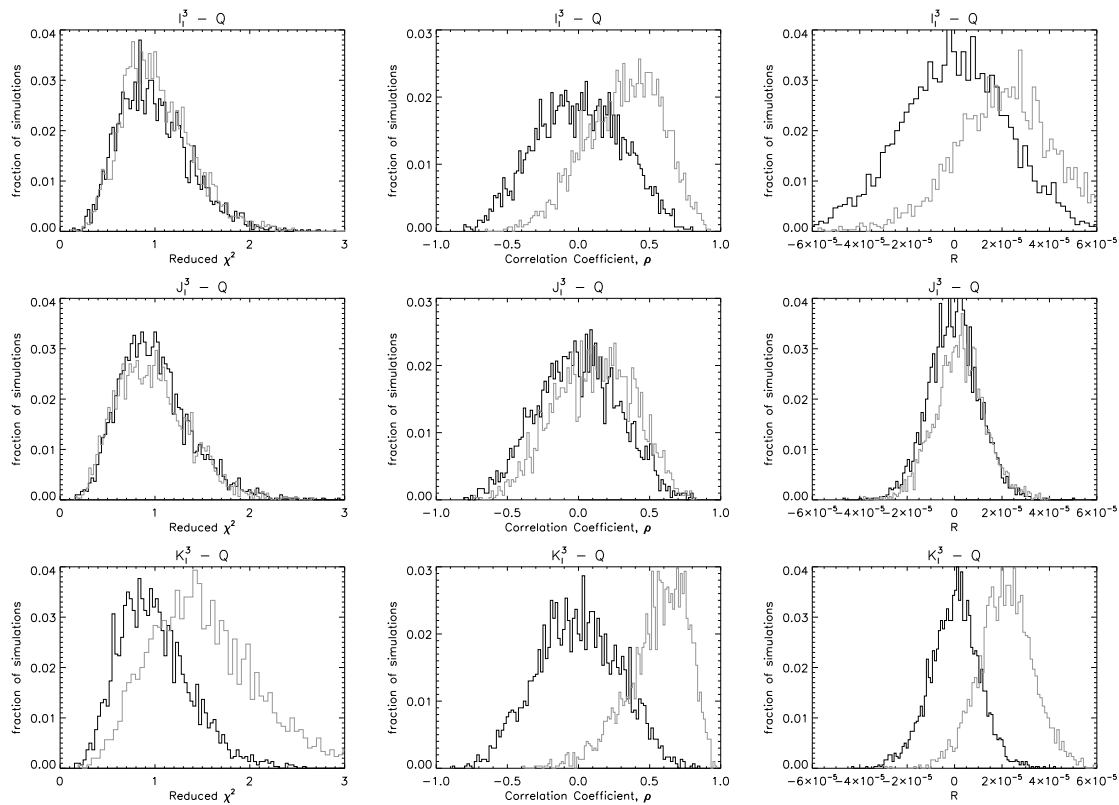,width=15cm}}
 \caption{ Distributions of values obtained for the three different
 tests applied to the Q channel to detect the presence of foregrounds
 ($\chi^2$, $ \rho $ and ${\rm R}$), for Gaussian simulations (black)
 and contaminated ($\alpha=1.0$) simulations (grey). The level of
 sensitivity of a given method can be determined in terms of the shift
 between the histograms for the Gaussian and the contaminated case. We
 see that the $\rho^{K}$ and ${\rm R}^K$ statistics are the most
 sensitive channels to probe the existence of foregrounds.}
 \label{fig:histo}
\end{figure*}

Again we test the sensitivity of this method by computing a
distribution of ${\rm R}$ for Gaussian ensembles against the
contaminated ensembles. We make sure that for Gaussian ensembles we
use the correlation of $\hat C^{\rm S+F}_\ell$ with $X^{\rm S}_\ell$
and for the contaminated ensemble the correlation of $\hat C^{\rm
S+F}_\ell$ with $X^{\rm S+F}_\ell$ where $\rm S$ stands for the
Gaussian CMB signal and $\rm S+F$ indicates contaminated
ensembles. This allows us to cancel the effect of the increase of
power due to foregrounds in the correlation of the two statistics
between the two tests. The results for the contaminated ensemble,
$\alpha=1$, are plotted in the right column of Fig.~\ref{fig:histo}
and are summarized in table~\ref{tab:corr} for all three bands.

\begin{table}
  \caption{ Sensitivity of the $\rho$ and ${\rm R}$ tests in terms of
  the fraction of the contaminated simulations ($\alpha=1.0$) with a
  larger value than 95.45 $\%$, ie 2 $\sigma$, of the Gaussian
  simulations ($\alpha =1.0$).  We present values for the Q,V and W
  frequency channels and for the $I^3_\ell$, $J^3_\ell$ and
  $K^3_\ell$. Note that $\rho(X_\ell)$, where $X_\ell$ is a given
  bispectrum component stands for the correlation between $X_\ell(\rm
  data)$ with$X_\ell(\rm template)$, whereas ${\rm R}(X_\ell)$,
  represent the correlation of that specific bispectrum component with
  the respective power spectrum of the map.  The values in the table
  quantify what can be seen in the histograms in
  figure~\ref{fig:histo}. Applying the tests for the $K^3_\ell$
  component provides better sensitivity to the foregrounds. Between
  the two tests $\rho$ seems to provide a marginally better
  sensitivity. Also, the Q channel, being the most foreground
  contaminated yields the higher chances of detection.}
 \begin{tabular}[b]{|c|c|c|c||c|c|c}
\hline & \multicolumn{1}{|c|}{ $\rho_{\rm Q}$ } &
 \multicolumn{1}{|c|}{ $\rho_{\rm V}$ } & \multicolumn{1}{|c|}{
 $\rho_{\rm W}$ } & \multicolumn{1}{|c|}{ ${\rm R}_{\rm Q}$ } &
 \multicolumn{1}{|c|}{ ${\rm R}_{\rm V}$} & \multicolumn{1}{|c|}{
 ${\rm R}_{\rm W}$ } \\
\hline\hline
$I^3_\ell$ & 0.541& 0.085 & 0.139 & 0.280& 0.030 & 0.080 \\
$J^3_\ell$ & 0.225& 0.100 & 0.091 & 0.080& 0.060 & 0.060 \\
$K^3_\ell$ & 0.714& 0.072& 0.113& 0.690& 0.290 & 0.110 \\
 \hline
\end{tabular}
\label{tab:corr}
\end{table}

\begin{table}
  \caption{ Results for the WMAP data for $\rho$ and ${\rm R}$. The
  results are shown as the fraction of Gaussian simulations below the
  level obseverved  in the data. We have highlighted values with
  greater than 98\% in the foreground cleaned maps.}
 \begin{tabular}[b]{|c|c|c|c|c|c|c}
 \hline & \multicolumn{3}{|c|}{ RAW } & \multicolumn{3}{|c|}{
 CLEANED}\\
\hline & \multicolumn{1}{|c|}{
 ${\rm Q}$ } & \multicolumn{1}{|c|}{ ${\rm V}$} &
 \multicolumn{1}{|c|}{ ${\rm W}$ }& \multicolumn{1}{|c|}{ ${\rm Q}$ }
 & \multicolumn{1}{|c|}{ ${\rm V}$} & \multicolumn{1}{|c|}{ ${\rm W}$
 }\\
\hline\hline
${\rm R}^I$ & 0.726 & 0.178 & 0.328 & 0.475 & 0.421 &        0.775  \\
${\rm R}^J$ & 0.758 & 0.802 & 0.749 & 0.822 & 0.869 & $\fbox{0.983}$\\
${\rm R}^K$ & 0.983 & 0.450 & 0.486 & 0.362 & 0.364 &        0.188 \\
\hline
$\rho^{I}$ & 0.998 & 0.408 & 0.762 &        0.491&          0.550& 0.452\\
$\rho^{J}$ & 0.933 & 0.906 & 0.856 & $\fbox{0.998}$& $\fbox{0.985}$ & $\fbox{0.986}$\\
$\rho^{K}$ & 0.922 & 0.166 & 0.272 &        0.013 &         0.021 & 0.044 \\
\hline
\end{tabular}
\label{tab:data}
\end{table}

\section{Application to the WMAP data}\label{sec:application}

We have applied the statistical tools described above to the WMAP
first year data \citep{wmap}. We considered both the raw and
cleaned maps of the Q, V, and W channels
using the Kp2 exclusion mask. We summarise the results in
table~\ref{tab:data} showing the separate confidence limits from
each channel for both the raw and cleaned maps.

For the raw maps we find that only the Q channel ${\rm R}^K$
result is above the 95\% threshold while for the Q channel $\rho$
statistic, all confidence levels are above the 90\% level with the $\rho^I$
above the 95\% level. This is consistent with there being a component
most correlated to the foreground templates at the lowest frequencies
and with significant correlations between the $\Delta\ell= 2$ inter-$\ell$
bispectrum and the power spectrum. Since the raw maps do not have any
foreground subtracted from them this is not a surprise although the
confidence level suggests that the correlations are larger than what
was found for the expected amplitude ($\alpha=1$) of the foregrounds.

For all $I^3_\ell$ and $K^3_\ell$ statistics the cleaned map results
show confidence levels below the 95\% level and indeed show an overall
reduction in the significance of the correlations, indicating that the
cleaning has removed a component correlated to the foreground
templates, as one would expect. However for the $J^3_\ell$ statistics,
which should in principle be the least sensitive to the foregrounds
considered, we see that the confidence levels have all
increased. Indeed all three channels now have correlations significant
above the 95\% level in the $\rho$ statistic with the W channel also
having a~$>95\%$ confidence level. The cleaning algorithm appears to
have introduced significant correlations with the foreground templates
in the $\Delta\ell=1$ inter-$\ell$ bispectra and significant correlations
between the $\Delta\ell=1$ inter-$\ell$ bispectrum and power spectrum of
the W channel which is indicative of a non-Gaussian component.

\begin{figure*}
\centerline{\psfig{file=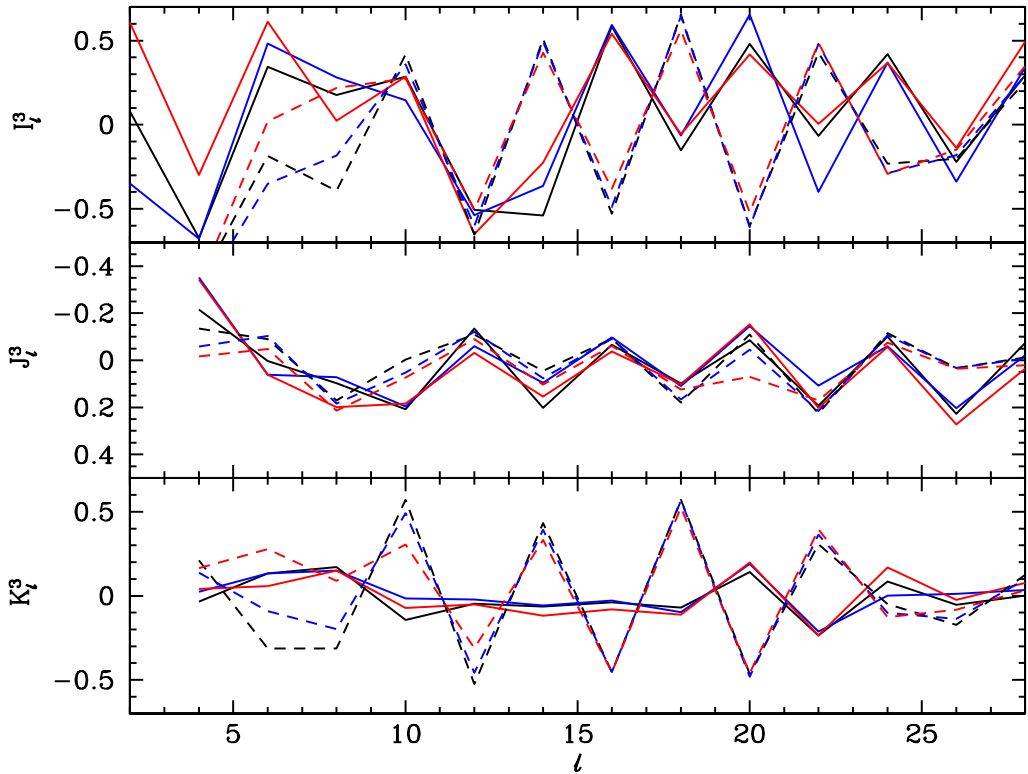,angle=270,width=15cm}}
 \caption{ The bispectra for each cleaned channel map (solid line)
against the bispectra of the foreground template ($\alpha=1.0$)
(dashed lines). The three frequency channels are shown as Q (red), V
(black) and W (blue). Both the cleaned map $I^3_\ell$ and
$K^3_\ell$ bispectra are anti-correlated with the foreground
templates. In addition the the $K^3_\ell$ for all channels are heavily
suppressed in the cleaned maps for multipoles $\ell < 20$ compared to
the expected Gaussian variance shown in figure~\ref{fig:bisp}.  The
$J^3_\ell$ gives the only bispectra that are correlated with the those
of the templates.} \label{fig:bisp_data}
\end{figure*}

In figure~\ref{fig:bisp_data} we show the bispectra for each cleaned
channel map and compare to the bispectra of the foreground
template ($\alpha=1$) for each channel. This shows the nature of
the result above. For both the $I^3_\ell$ and $K^3_\ell$ the
cleaned map bispectra are anti-correlated with the foreground
templates. In addition the the $K^3_\ell$ for all channels are
heavily suppressed in the cleaned maps for multipoles $\ell < 20$
compared to the expected Gaussian variance shown in
figure~\ref{fig:bisp}. The $J^3_\ell$ gives the only bispectra that
are correlated with the those of the templates.

\begin{figure*}
\centerline{\psfig{file=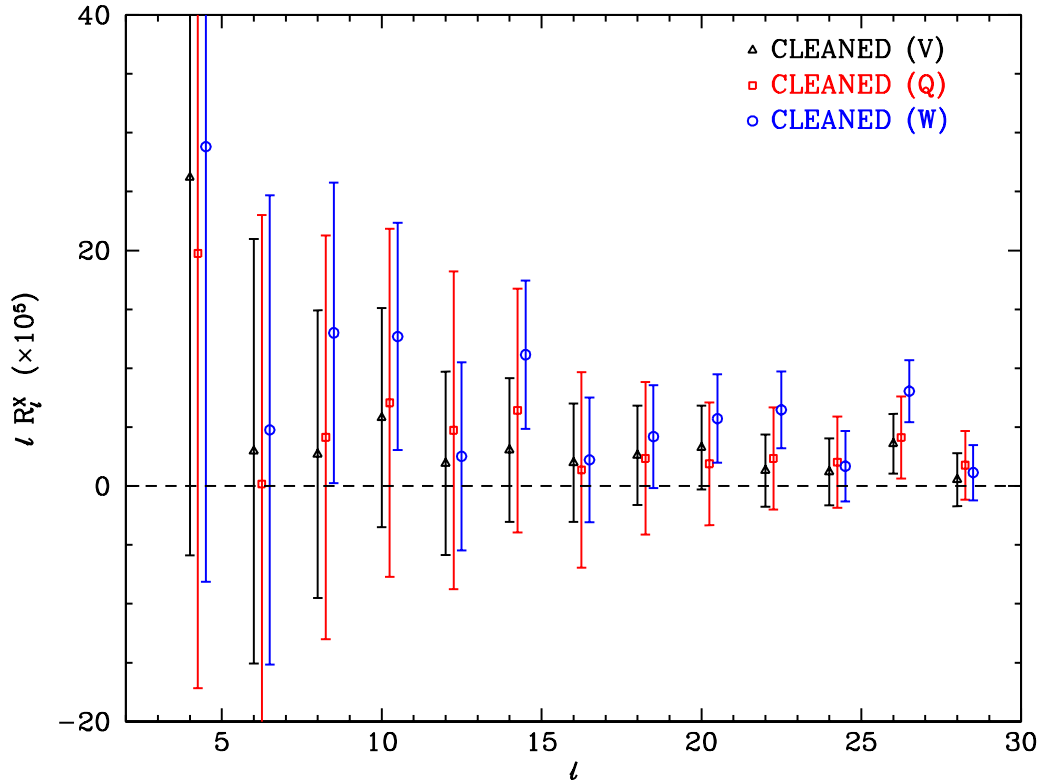,angle=270,width=15cm}}
 \caption{ ${\rm R}^J$ as a function of angular scale, $\ell$. The
values displayed correspond to the foreground cleaned Q (red squares),
V (black triangles) and W (blue circles) frequency channels. The
values are offset by $\ell=0.25$ and $\ell=0.5$ for the Q and W bands
respectively. The $3 \sigma$ detection in the W foreground-cleaned
channel is dominated mainly by the $\ell=26$ mode. The error bars are
computed from 3000 Gaussian simulations assuming the specific channel
noise and beam.}
\label{fig:data}
\end{figure*}

Figure~\ref{fig:data} shows the break down of the $R^J$ result into
individual multipole contributions for each of the three bands. In
particular it is interesting to note how the W band ${\rm R}^J$ result
shown in table~\ref{tab:data} is dominated by an outlier at $\ell=26$.

\section{Discussion}\label{sec:disc}

At first sight our results appear contradictory. We have studied the
effect of foreground contamination on the maps and concluded that
foregrounds mainly affect the $I^3_\ell$ and $K^3_\ell$ components
of the bispectrum due to its parity. By comparing  the results for
the raw and the foreground-cleaned maps, we are able to verify that
the amplitude of $I^3_\ell$ and $K^3_\ell$ reduces as expected after
foreground subtraction.

On the other hand, as shown in table~\ref{tab:data}, the correlations
induced in the $J^3_\ell$ appear to be close to inconsistent to a
Gaussian hypothesis with the correlation with the foreground templates
at a significance above the $3\sigma$ level for the Q-band, cleaned
map. It is also of interest to note that the cleaned maps do worse in
all bands for the $\rho$ measure.

This is not what we naively expected since the foregrounds considered
here have the wrong parity and their $J^3_\ell$ signal is heavily
suppressed. However the cleaning procedure used by the WMAP team {\sl
does} appear to increase the correlations $\rho$ of $J^3_\ell$
bispectrum to the input maps and its correlation $\rm R$ with the
power spectrum. Recall that we expect the normalized bispectra to be
independent of the power spectrum only in the Gaussian case.

The possibility of the foregrounds being more complex than accounted
for in this type of treatment is to be considered carefully as this
work has shown. The results shown here would suggest that the
procedure used to go from the {\sl raw} to {\sl cleaned} WMAP maps is
under or over subtracting a component with $\ell\pm 1$ parity in the
bispectrum. This is probably not an indication that the procedure is
faulty but rather that the templates used are not accurate enough to
subtract the foregrounds. One source of inaccuracy is the simple
scaling of the templates with respect to frequency. The {\sl cleaned}
maps are obtained assuming uniform spectral index and \cite{wmapfor}
acknowledge that this is a bad approximation particularly for the 408
MHz Haslam (synchrotron) template. This is seen when producing the
Internal Linear Combination (ILC) map which accounts for variation of
the spectral index of the various component. Unfortunately ILC maps
cannot be used in quantitative studies as their noise attributes are
complicated by the fitting procedure and one cannot simulate them
accurately. 

Future WMAP ILC maps or equivalent ones obtained by
`blind' foreground subtraction \citep{tegclean,erik2} may be
better suited for this kind of analysis once their statistical
properties are well determined. It is expected that the impending
second release of WMAP data will allow more accurate foreground
analysis and the statistical tools outlined in this work will be useful in
determining the success of foreground subtraction.

It may be worthwile to include information of the higher order
statistics when carrying out the foreground subtraction itself, for
example by extending the ILC method to minimise higher order map
quantities such as the skewness and kurtosis of the maps.

\section*{Acknowledgments}
We thank  H. K. Eriksen for advice and for making the
simulations available to us. We are also grateful to Jo\~ao
Magueijo, Kate Land and A.J. Banday for useful conversations
throughout the preparation of this work. Some of the results in
this paper have been derived using the HEALPix package. J.
Medeiros acknowledges the financial support of Fundacao para a
Ciencia e Tecnologia (Portugal).

\end{document}